\documentclass[reprint,floatfix,
 amsmath,amssymb,
 aps,
showkeys,
prb
]{revtex4-2}
\usepackage{booktabs}
\usepackage{graphicx}
\usepackage{dcolumn}
\usepackage{bm}
\usepackage[section]{placeins}
\usepackage{multirow}
\usepackage{dirtytalk}
\usepackage{natbib}
\usepackage{xcolor}
\usepackage{hyperref}
\hypersetup{
    colorlinks=true,       
    linkcolor=black,        
    citecolor=blue,        
    filecolor=blue,     
    urlcolor=black          
}

\usepackage{xr} 
\newcommand{\supplementarysection}{%
  \setcounter{section}{0}
  \let\oldthesection\thesection
  \setcounter{figure}{0}
  \let\oldthefigure\thefigure
  \renewcommand{\thefigure}{S\oldthefigure}
  \setcounter{table}{0}
  \let\oldthetable\thetable
  \renewcommand{\thetable}{S\oldthetable}
  \section{Supplementary section}
}

\begin{document}

\title{Visualizing Anisotropic Charge Transport in CrSBr using Bias-Dependent Kelvin Probe Force Microscopy}

\author{Satyam Sahu$^1$}
\email{satyam.sahu@heu.cas.cz}
\author{Arijit Kayal$^1$}
\email{arijit.kayal@heu.cas.cz}
\author{Bing Wu$^2$}%
\author{Aljoscha Söll$^2$}%
\author{Zdeněk Sofer$^2$}
\author{Matěj Velický$^1$}%
\author{Otakar Frank$^1$}%
\email{otakar.frank@heu.cas.cz}
\affiliation{%
 $^1$Heyrovský Institute of the Czech Academy of Sciences, Dolejškova 2155/3, 182 00, Prague 8, Czech Republic\\$^2$Department of Inorganic Chemistry, University of Chemistry and Technology Prague, Technická 5, 166 28 Prague 6, Czech Republic}%

\date{\today}

\begin{abstract}
Anisotropic semiconductors offer unique opportunities for next-generation electronics due to their direction-dependent transport. However, conventional electrical measurements provide only spatially averaged characteristics. Here, we employ bias-dependent Kelvin probe force microscopy (KPFM) to directly visualize the response of CrSBr under different electrical bias configurations. Spatially resolved KPFM maps reveal highly distinct, crystallographic-direction-dependent potential profiles. To elucidate this behavior, we employ anisotropic transport simulations that capture the charge transport and electric-field distributions in CrSBr. The close agreement between experiment and simulation provides a unified picture of anisotropic charge transport and bias-induced potential redistribution. Our experiments further reveal a boundary-associated redistribution of the transport field when bias is applied along the $a$-axis, suggesting a possible role of crystal boundaries in charge transport along the low-conductivity direction. This work establishes bias-dependent KPFM as a powerful platform for uncovering localized transport phenomena in low-symmetry quantum materials.

\end{abstract}

\maketitle

\section{Introduction}
Low-dimensional materials with strong in-plane anisotropy exhibit direction-dependent charge transport and other physical properties governed by their crystallographic symmetry \cite{Klein2023, Wu2022, Telford2023, Volochanskyi2026, Ye2026, Li2021}. Their directional electronic response provides opportunities for anisotropic electronics \cite{Wu2022, Panda2023, Volochanskyi2026, Zhao2025, Khan2024}. Understanding how an applied electrical bias is distributed across such anisotropic materials is therefore essential for both fundamental studies and the development of next-generation electronic devices.\par

Among recently discovered layered semiconductors, CrSBr has attracted considerable attention owing to its unique combination of strong in-plane anisotropy, intrinsic magnetic ordering, and environmental stability \cite{Klein2023, Ziebel2024, Telford2023}. Its orthorhombic crystal structure gives rise to highly anisotropic electronic and optical properties \cite{Wu2022, Sahu2025, Linhart2023}, making CrSBr a model system for studying anisotropic charge transport and spin-optoelectronic phenomena \cite{Dong2026, Chen2024, Ye2022}. Electrical transport measurements have consistently demonstrated pronounced differences between the crystallographic $a$- and $b$-axes, highlighting the strong directional character of charge transport with a conductivity ratio ($\sigma_b/\sigma_a$) of $10^2 - 10^5$ (at 10 K) in this material \cite{Wu2022}. However, understanding how this anisotropy manifests spatially within a device requires access to the local potential and electric-field distribution under applied bias.\par

Kelvin probe force microscopy (KPFM), which measures the contact potential difference (CPD) between a conductive probe and the sample, provides nanoscale access to spatial variations in surface potential and work function \cite{Melitz2011, Zahmatkeshsaredorahi2025}. KPFM has been widely used to investigate electronic properties of semiconductor devices and interfaces \cite{Zahmatkeshsaredorahi2025}. In an electrically biased device, spatial variations in the measured potential can be used to reconstruct equipotential contours and, consequently, the local electric-field distribution \cite{Dagan2019, Kayal2021}. However, bias-dependent KPFM studies that connect the measured potential landscape to the local electric-field distribution in intrinsically anisotropic layered semiconductors remain comparatively unexplored.\par

Here, we employ bias-dependent KPFM to map the spatial evolution of the local potential in CrSBr under an applied electrical bias along its two principal axes. From the measured potential maps, we extract equipotential contours and local electric fields, enabling a spatially resolved comparison of the electrical response along the $a$- and $b$-axes. We further compare the experimentally obtained potential and electric-field distributions with anisotropic transport simulations to establish the relationship between conductivity, potential redistribution, and electric-field concentration. The comparison reveals distinct transport characteristics in the two directions and suggests an additional boundary-associated potential-redistribution mechanism when the bias is applied along the low-conductivity $a$-axis. Together, these results demonstrate how bias-dependent KPFM, combined with numerical modeling, can provide insight into anisotropic charge transport and reveal transport pathways not apparent from conventional measurements.
\section{Experimental Details}\label{sec:methods}
We synthesized bulk CrSBr single crystals by chemical vapor transport \cite{Sahu2026}. We first patterned device substrates consisting of highly doped Si with a 285 nm thermally grown SiO$_2$ layer using standard optical lithography. We then deposited Cr/Au (5/35 nm) electrodes by magnetron sputtering, followed by a lift-off process to define the contact geometry. CrSBr flakes were mechanically exfoliated from the bulk crystals using conventional adhesive tape (Nitto Denko Corp.) and deterministically transferred onto the pre-patterned electrodes under an optical microscope using a standard PDMS (polydimethylsiloxane) stamp-assisted dry-transfer method to fabricate the devices  \cite{Panda2023, Sahu2025_2}. For the top-gated devices, an hBN flake was subsequently transferred onto the CrSBr flake, followed by the transfer of a few-layer graphene flake to form the top gate.\par
We characterized the thickness and surface morphology of the transferred CrSBr flakes using atomic force microscopy (AFM). The crystallographic orientation of each flake was determined by polarization-resolved Raman spectroscopy using a 532 nm excitation laser and an 1800 lines/mm grating in backscattering geometry. We used the characteristic polarization dependence of the Raman-active phonon modes to identify the crystallographic $a$- and $b$-axes (Supplementary Figure S1). These orientations were subsequently used as the reference directions for all electrical transport and KPFM measurements.\par
Direction-dependent electrical transport ($I_{\mathrm{ds}}$--$V_{\mathrm{ds}}$) characteristics were measured under ambient conditions using a Keithley 2612B SourceMeter.\par
We performed the KPFM measurements using a HORIBA OmegaScope scanning probe microscope (Horiba Jobin-Yvon) operating in frequency-modulation (FM-KPFM) mode with Au-coated conductive probes (ACCESS-FM-GG, Applied NanoStructures Inc.) \cite{Sahu2025_2}. The surface potential was recorded under different $V_{\mathrm{ds}}$ applied to the pair of electrodes defined along the $a$- and $b$-axes of the CrSBr flake. The electrical bias was supplied using the Keithley 2612B SourceMeter, while the KPFM tip monitored the local CPD, enabling spatial mapping of the local surface potential across the CrSBr device. Zero-bias KPFM scans were performed after every electrode-configuration change to verify the tip stability and establish a common background for non-zero-bias measurements.
\section{Results and Discussion}
\begin{figure}[htbp!]
    \centering
    \includegraphics[width=3.3in]{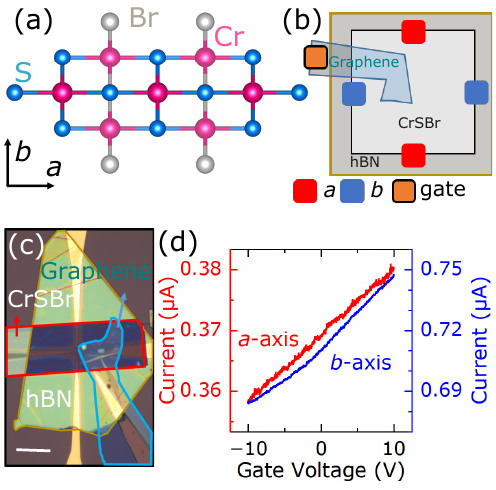}
    \caption{(a) Schematic of the orthorhombic crystal structure of CrSBr. (b--c) Schematic illustration and magnified optical image of a top-gate CrSBr field-effect transistor. Scale bar: 5 {\textmu}m. (d) Transfer characteristics of the top-gate CrSBr device when the $V_{\mathrm{ds}}$ is applied along the $a$-axis (red) and the $b$-axis (blue).}
    \label{fig:char}
\end{figure}
As shown in Figure \ref{fig:char}a, CrSBr crystallizes in an orthorhombic layered structure with $Pmmn$ symmetry and pronounced in-plane anisotropy. The two inequivalent crystallographic axes ($a$ and $b$) give rise to a strongly direction-dependent structural and electronic response \cite{Klein2023, Wu2022, Sahu2025}. Consistent with this anisotropy, mechanically exfoliated CrSBr flakes typically exhibit an elongated, rod-like morphology, with the longer and shorter dimensions corresponding to the $a$- and $b$-axes, respectively. This anisotropy is associated with the quasi-1D electronic character of CrSBr, with charge transport preferentially favored along the $b$-axis \cite{Klein2023, Wu2022}.\par
\begin{figure*}[htbp!]
    \centering
    \includegraphics[width=6.2in]{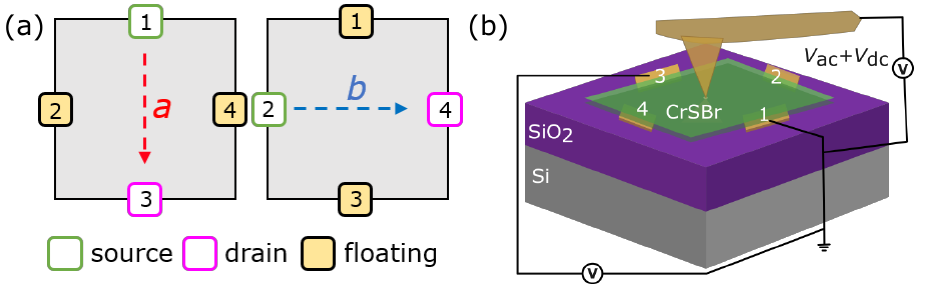}
    \caption{(a) Measurement configurations for directional profiling along the two axes. (b) Schematic layout of the KPFM setup with $V_{\mathrm{ds}}$ applied to the CrSBr channel according to the configurations illustrated in panel (a).}
    \label{fig:scheme}
\end{figure*}
To understand the directional transport response, we performed gate-dependent electrical measurements on CrSBr devices incorporating a graphene/hBN/CrSBr top-gate structure, as schematically illustrated in Figure \ref{fig:char}b. Figure \ref{fig:char}c shows an optical image of a representative device, highlighting the CrSBr flake, graphene gate, and hBN dielectric. Figure \ref {fig:char}d  and Supplementary Figure S2 trace the source-drain current ($I_{\mathrm{ds}}$) as a function of top-gate voltage ranging from -10 V to +10 V at a fixed source-drain voltage $V_\textrm{ds}$ = 0.1 V for transport along $a$- and $b$-axes. For both orientations, the current increases monotonically with an increase in the gate voltage, yielding transconductance slopes of 1.0 nA/V and 3.7 nA/V for the bias along the $a$-axis and the $b$-axis, respectively. The positive transconductance is consistent with n-type field-effect behavior, in agreement with a recent report \cite{Biktagirov2025}.\par
From this directional transport response, we extracted the apparent two-terminal field-effect mobility values. The mobilities obtained were 0.2 and 0.7 cm$^2$ V$^{-1}$ s$^{-1}$ along the $a$-axis and the $b$-axis, respectively, revealing a distinct apparent mobility anisotropy ($\mu_b/\mu_a$) of $\approx$ 3.5. Assuming comparable carrier densities in both directions, this ratio also estimates a similar conductivity anisotropy in CrSBr. The detailed calculations are provided in Supplementary Information Section S2.\par

Figure \ref{fig:scheme}a illustrates the two bias configurations used to investigate charge transport in CrSBr. $V_\textrm{ds}$ is applied along either the $a$- or $b$-axis, while the remaining contacts are left electrically floating. To assess the reproducibility of the directional transport response, we measured room-temperature $I_{\mathrm{ds}}-V_{\mathrm{ds}}$ characteristics along both axes for six devices. As shown in Supplementary Figure S3, all devices exhibit higher $I_{\mathrm{ds}}$ along the $b$-axis than along the $a$-axis, consistent with the higher $b$-axis mobility extracted from the top-gate measurements. Nevertheless, the absolute current varies between devices, reflecting the channel geometry dependence of the two-terminal current. These measurements establish the reproducibility of the directional trend.\par
While directional $I_{\mathrm{ds}}$--$V_{\mathrm{ds}}$ measurements capture the macroscopic anisotropy of charge transport, they provide an averaged device response and cannot resolve the local potential landscape within the channel. In an ideal uniform channel with ohmic contacts, the potential is expected to vary approximately linearly between the source and drain. However, in real devices, contact resistance and local structural or topographical features, such as wrinkles, step edges, and polymer residues, can modify the local potential distribution and produce additional potential drops. These spatial variations cannot be resolved from the global $I_{\mathrm{ds}}$--$V_{\mathrm{ds}}$ characteristics alone. Furthermore, macroscopic transport cannot reveal how the local potential distribution responds to bias applied along different axes or establish a relationship between the local electric field and transport anisotropy.\par
To address this limitation, we employed bias-dependent KPFM, using the measurement geometry illustrated in Figure \ref{fig:scheme}b. By mapping the CPD while applying $V_\textrm{ds}$, KPFM enables spatially resolved visualization of the local electrical potential response across the CrSBr channel. This approach allows us to examine how the potential landscape evolves with bias direction and to identify spatial variations obscured in macroscopic electrical measurements. 

\begin{figure*}[htbp!]
    \centering
    \includegraphics[width=7.0in]{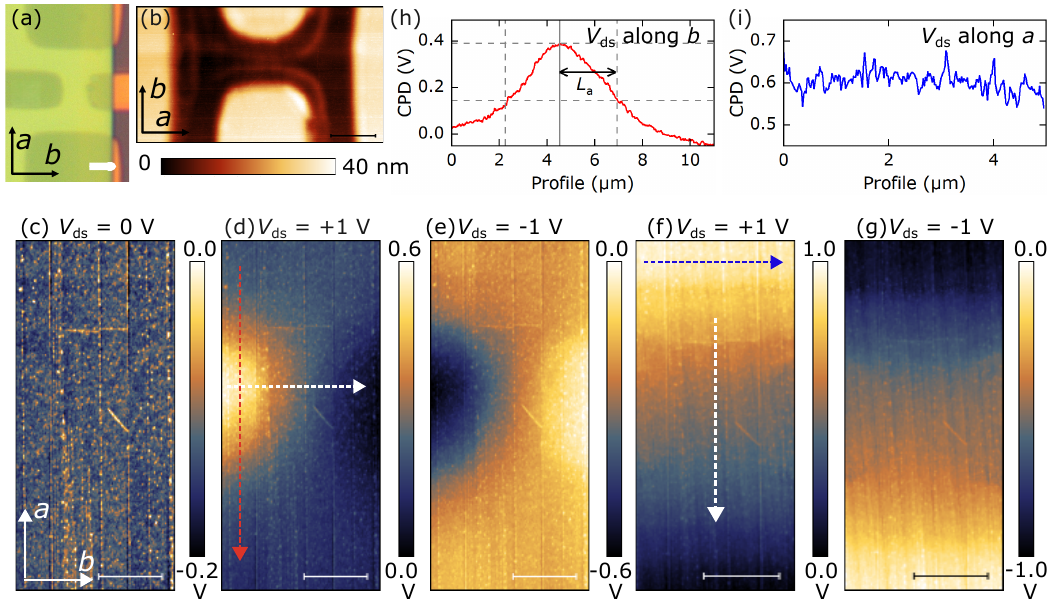}
    \caption{ (a) Optical and (b) AFM images of the investigated CrSBr (thickness: $40$ nm) device with four Cr/Au contacts. KPFM CPD maps under (c) $V_\textrm{ds}$ = 0.0 V and $V_\textrm{ds}$ = $\pm$1.0 V along (d--e) the high-conductivity $b$-axis and (f--g) the low-conductivity $a$-axis. All scale bars: 2 {\textmu}m. (h) and (i) Transverse line profiles for bias along the $b$- and $a$-axes along the same-colored arrows shown in (d) and (f), respectively.}
    \label{fig:cpd}
\end{figure*}
Figures \ref{fig:cpd}a--b show the optical and AFM images of one of the devices investigated in this work. This device has asymmetric channel lengths along the two measurement configurations. To establish that the observed response is not unique to this geometry, we also measured a device with symmetric channel lengths and observed qualitatively identical behavior (Supplementary Figure S4). The following analysis therefore focuses on the spatially resolved potential response in the representative asymmetric device, with the symmetric device measurements supporting the robustness of the observation.\par
Figure \ref{fig:cpd}c shows the CPD map acquired at $V_{\mathrm{ds}}=0$ V. In the absence of an externally applied bias, the CrSBr channel exhibits no systematic source–drain potential gradient, with superimposed local CPD variations, attributed to wrinkles, surface adsorbates, and defects that generate electrostatic inhomogeneities \cite{Torres2023, Sovizi2026}. The absence of a noticeable CPD gradient suggests that the Fermi level remains flat across the entire channel. The CPD varies between $0$ V and $-0.2$ V, with an average of $\approx-0.1\pm0.02$~V. Importantly, this zero-bias map provides the static potential background under equilibrium conditions against which the subsequent bias-induced potential redistribution can be distinguished. \par

Upon applying $V_{\mathrm{ds}}=\pm1$ V along the high-conductivity $b$-axis, a pronounced lateral CPD gradient develops across the channel (Figures \ref{fig:cpd}d--e), reflecting a steady-state current flowing through the device. Switching the polarity of the $V_{\mathrm{ds}}$ reverses the direction of the potential gradient, demonstrating that the observed contrast is predominantly associated with the externally applied bias and the contacts' symmetry. Both CPD maps in Figures \ref{fig:cpd}d--e show a well-defined structure of the underlying Au electrodes, with one of them at $V_{\mathrm{ds}}=\pm1$ V and the other grounded. Similar behavior involving a linear potential drop and bias reversal was observed when $V_{\mathrm{ds}}$ was applied along the low-conductivity $a$-axis, as shown in Figures~\ref{fig:cpd}f–-g. However, the potential variation extends over a substantially larger distance because the channel is longer along the $a$-axis. CPD maps acquired at intermediate bias voltages are provided in Supplementary Figure S5.\par
To understand these spatial variations, we extracted CPD line profiles along the respective channel directions (Supplementary Figure S5). The profiles reveal an approximately linear potential drop across the channel for non-zero bias, with local deviations superimposed on the overall gradient. The corresponding macroscopic electric fields, estimated from the potential gradients, reverse sign with the applied bias polarity.\par
Notably, the CPD distribution in Figures~\ref{fig:cpd}d--g surrounding the contact pads differs between the two cases. For transport along the $b$-axis (Figures~\ref{fig:cpd}d--e), the CPD is more localized near the Au electrodes and drops sharply, following the underlying electrode geometry. In contrast, when the bias is applied along the $a$-axis (Figures~\ref{fig:cpd}f--g), the CPD appears highly diffused and spreads over larger spatial length scales. This behavior is reproduced in the CPD maps for the symmetric channel-length device shown in Supplementary Figure S4. The asymmetry in the CPD profiles for transport along the two axes stems from the inherent anisotropy of CrSBr.\par
To quantify the spatial extent of the potential redistribution, we extracted transverse line profiles from the CPD maps, perpendicular to the direction of $V_\textrm{ds}$ (= 1 V) as shown in Figures~\ref{fig:cpd}h--i. These profiles correspond to the CPD maps in Figures ~\ref{fig:cpd}d and f, respectively. For $b$-axis transport, the transverse potential profile in Figure \ref{fig:cpd}h exhibits a characteristic decay away from the region of maximum potential, yielding an effective transverse potential-spreading width ($L_\textrm{a}$) $\sim$ 2.4 {\textmu}m using the $V = V_\textrm{max}/e$ criterion, i.e., the distance from the maximum potential position to the point where the potential decreased to 37\%. Here, the subscript $a$ indicates that the extracted diffusion occurs along the $a$-axis, transverse to the applied bias. For transport along the $a$-axis, the corresponding effective transverse potential-spreading width along the $b$-axis is provisionally estimated by applying a square-root diffusivity scaling, using $L_\textrm{b}/L_\textrm{a} = \sqrt{D_\textrm{b}/D_\textrm{a}}$, yielding $L_b^{\textrm{pred,}\sqrt{D}}\approx 4.5$ {\textmu}m (see Supplementary Information Section S2 for details), as it is not possible to measure it experimentally from our data due to the flatness of the potential in the measured range (Figure \ref{fig:cpd}i). This estimate will be examined below using anisotropic-conduction simulations.\par
\begin{figure*}[htbp!]
    \centering
    \includegraphics[width=7in]{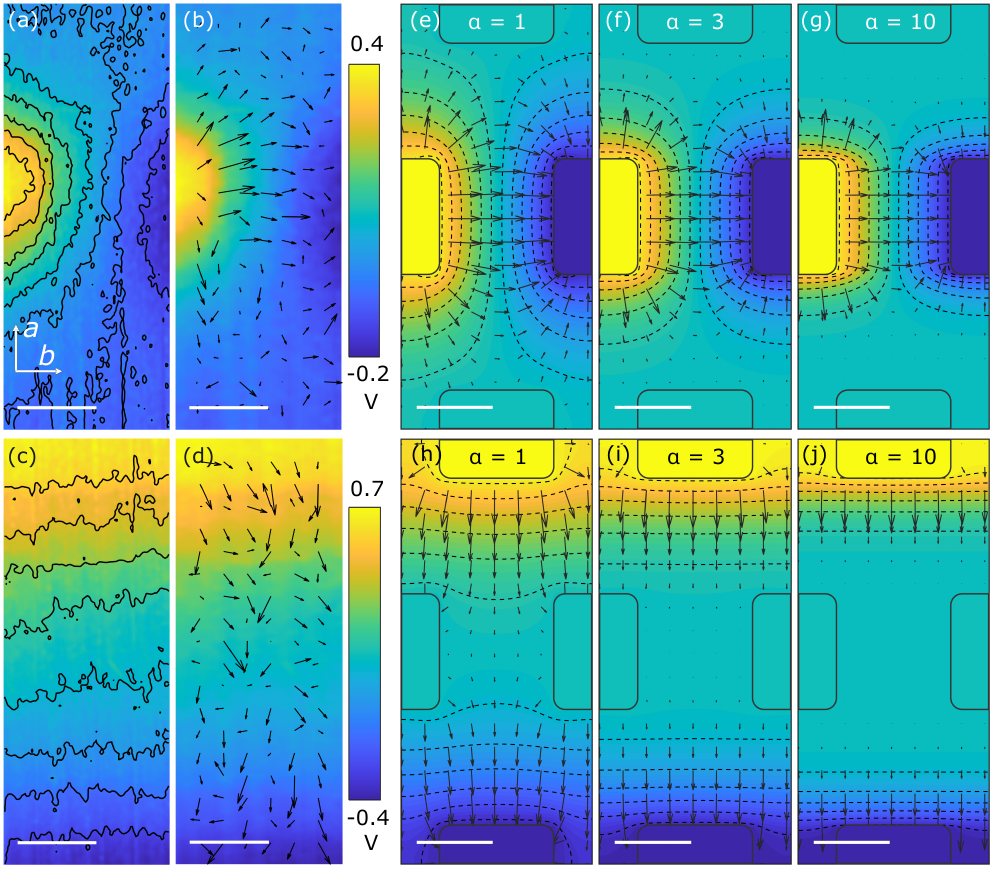}
    \caption{Experimental CPD maps with equipotential contour lines for $V_\textrm{ds} = +1$~V along: (a) high-conductivity $b$-axis, and (c) low-conductivity $a$-axis. Vector maps of local electric field lines calculated from CPD maps for $V_\textrm{ds} = +1$~V along: (b) $b$-axis, and (d) $a$-axis. Vectorized numerical simulations modeling potential contours and electric field lines for bias along (e--g) $b$-axis, and (h--j) $a$-axis for three different conductivity anisotropy ratios ($\alpha=\sigma_b/\sigma_a$): isotropic ($\alpha = 1$), moderate anisotropy similar to the experimental value from the mobility ratio ($\alpha = 3$), and strong tensor confinement ($\alpha = 10$).  All scale bars: 2 {\textmu}m.}
    \label{fig:contours}
\end{figure*}

While electrical measurements establish that conduction is more efficient along the $b$-axis, the KPFM data directly demonstrate how this directional anisotropy manifests in the spatial redistribution of the applied potential. To further resolve these differences, we reconstruct the two-dimensional equipotential landscapes and calculate the corresponding local electric-field distributions from the measured CPD maps. We then compare these experimentally derived distributions with anisotropic charge-transport-based simulations.
\par
Figures \ref{fig:contours}a and \ref{fig:contours}c show the experimental CPD maps for $V_{\mathrm{ds}}=+1$ V along the high-conductivity $b$-axis and low-conductivity $a$-axis, respectively, with equipotential contours (also see Supplementary Figure S6). Figures \ref{fig:contours}b and \ref{fig:contours}d show the corresponding reconstructed local electric field distributions overlaid on the CPD maps. We calculated the local electric field from the spatial gradient of the local potential ($\mathbf{E}=-\nabla V$), with the arrows representing the local field direction and their length proportional to the relative field strength.
\par
These experimental maps reveal a clear directional contrast in the potential and field distributions. For bias along the $b$-axis (Figures \ref{fig:contours}a--b), the equipotential contours are curved, following the shape of the source contact, and are closely spaced near the electrodes, indicating that the applied voltage spreads radially away from the contacts with a dominant potential gradient towards the drain contact. The local electric field vectors, related to the current density through $\mathbf{J}=\sigma\mathbf{E}$, are stronger between the source and drain contacts and predominantly point towards the macroscopic field direction. In contrast, $\mathbf{E}$ is weaker and more disordered in the other regions away from the source-drain contacts. The field distribution, therefore, indicates that the applied potential is accommodated predominantly within the source–drain region, with comparatively weaker transverse redistribution along the $a$-axis. By contrast, when bias is applied along the $a$-axis (Figures \ref{fig:contours}c--d), the equipotential contours appear flat and equally spaced throughout the channel.\par
Additionally, the external $V_{\mathrm{ds}}$ appears to spread across the transverse direction ($b$-axis) with a potential gradient along the macroscopic field direction. In both cases, the contours are not smooth and are often distorted, indicating local charge inhomogeneities that modify the local electric field and disrupt local transport. The anisotropy in potential and field distribution is also evident in a symmetric device with identical channel length and contact pads (Supplementary Figure S7).
\par
To better comprehend this anisotropic behavior, we performed two-dimensional numerical simulations using a finite-difference relaxation scheme. In the steady state, charge conservation requires $\nabla\cdot\mathbf{J}=0$. Assuming a homogeneous carrier concentration in the channel, the transport will be Ohmic in nature and described as $\mathbf{J}=\mathbf{J}_{drift}=-\boldsymbol{\sigma}\nabla V$, where $\boldsymbol{\sigma}$ is the in-plane conductivity tensor, and the potential distribution satisfies the anisotropic Laplace equation \cite{Lundstrom2002}. With the coordinate axes aligned with the principal crystallographic directions, the equation reduces to
\begin{equation}
\alpha\frac{\partial^2V}{\partial x^2}
+\frac{\partial^2V}{\partial y^2}=0,
\end{equation}

\noindent where $\alpha=\dfrac{\sigma_x}{\sigma_y}$ \cite{Pierret1987, Nye1985} is the conductivity anisotropy ratio. Here, the $x$-axis is aligned with the high-conductivity $b$-axis and the $y$-axis with the low-conductivity $a$-axis. The alignment with the principal axes eliminates the off-diagonal conductivity terms, consistent with their negligible contribution as indicated by Hall measurements \cite{Wu2022}, allowing the effect of the anisotropy to be captured by the single parameter $\alpha$. Details of the numerical implementation are provided in the Supplementary Information Section S8.
\par
Figures \ref{fig:contours}e--j show the simulated potential and electric-field distributions for $\alpha$ = 1, 3, and 10, for a bias along the two axes. For the isotropic case ($\alpha$ = 1), the potential distribution is governed primarily by the device geometry. Increasing the conductivity anisotropy progressively modifies both the equipotential contours and the electric-field distribution. For a bias along the $b$-axis, increasing $\alpha$ from 1 to 10 (Figures \ref{fig:contours}e--g) concentrates the equipotential contours closer to the electrode regions, which are otherwise spread along the transverse direction in the isotropic case, leaving a comparatively uniform central channel. As $\alpha$ increases, the local fields become more concentrated between the source-drain contacts, indicating that the simulated current density is confined to the central part of the channel, with less spread in the transverse direction compared to the isotropic case. This higher directionality of current conduction stems purely from the anisotropic nature of the channel conductivity. Conversely, when the bias is applied along the low-conductivity $a$-axis, the potential variation remains distributed over a substantially larger portion of the channel; furthermore, the equipotential contours that were curved near the contacts become flatter and more densely packed as $\alpha$ increases. Simulations for other $\alpha$ values and a symmetric device are presented in Supplementary Figures S8--S10. 
\par
The contrasting transverse potential distributions for $b$- and $a$-axis bias therefore arise from the anisotropic conductivity tensor and the corresponding boundary conditions of the device geometry. For $a$-axis bias, the strongly suppressed longitudinal conductivity forces a broader redistribution of the potential along the transverse high-conductivity $b$-axis, whereas $b$-axis transport confines the potential variation more strongly to the source–drain direction.
\par
In addition, we extended the simulation to sample dimensions that would allow the extraction of transverse potential-spreading widths in both principal-axis directions, while keeping all other simulation parameters the same as in the case of Figures \ref{fig:contours}e--j. In such remote-boundary calculations, the transverse potential-spreading ratio increases approximately linearly over the sampled anisotropy range (Supplementary Figure S11). Linear interpolation to $\alpha=3.5$ gives $L_b/L_a\approx1.9$, corresponding to $L_b^\textrm{pred,model}\approx 4.6$~\textmu m when referenced to the measured $L_a$. This is approximately 3\% larger than the square-root diffusivity benchmark of 4.5 \textmu m introduced earlier. The comparison indicates that the square-root estimate provides a reasonable numerical approximation at moderate anisotropy.\par
In general, the correspondence between the experimental and calculated landscapes demonstrates how conductivity anisotropy manifests in the local potential and electric-field distributions. These maps reveal spatial features not accessible from device-level transport measurements alone, motivating further examination of the transverse potential distribution.
\par
\begin{figure}[htbp!]
    \centering
    \includegraphics[width=3.3in]{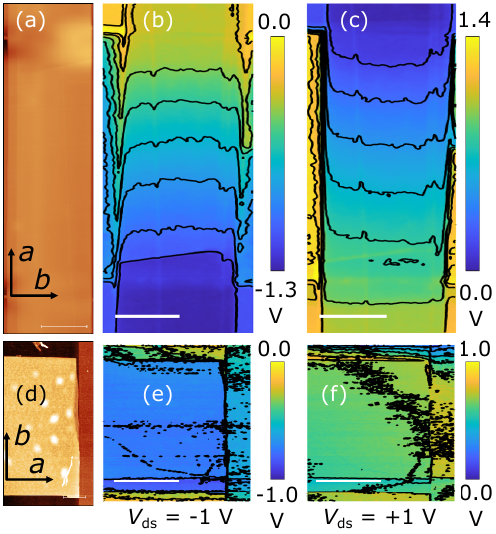}
    \caption{AFM images of the device and KPFM contact potential difference with equipotential contours at $V_\textrm{ds} = \pm1$~V for the bias along (a--c) the $a$-axis and (d--f) the $b$-axis. All scale bars: 2 {\textmu}m.}
    \label{fig:edge}
\end{figure}
Figure \ref{fig:edge} shows the AFM topography and CPD distributions under $V_{\mathrm{ds}}=\pm1$ V for CrSBr devices with two Au contacts configured to probe transport along the low-conductivity $a$-axis and the high-conductivity $b$-axis. For bias applied along the $a$-axis, the equipotential contours in Figures \ref{fig:edge}b--c exhibit a pronounced U-shaped profile across the channel, indicating a nonuniform potential drop across its width. The contours are more closely spaced near the lateral edges than in the central region, suggesting a stronger local electric field near the edges and a comparatively weak field in the channel interior. This spatial pattern persists upon bias reversal, with the potential gradient reversing direction, indicating that the edge-associated enhancement is linked to the applied bias rather than arising solely from a static potential inhomogeneity.\par
For comparison, we also examined the potential contours under bias applied along the $b$-axis (Figure \ref{fig:edge}e--f). Because of the flake's elongated morphology, the measurements capture only one lateral edge rather than the entire channel width. In this configuration, within the accessible $b$-axis field of view, the observed contours do not exhibit the U-shaped profile found for $a$-axis transport, showing that the potential drop is not concentrated at the lateral edges. Instead, the contrast is consistent with transport occurring predominantly through the central channel along the $b$-axis. \par
The contrasting potential distributions suggest that boundary-associated transport contributions are more prominent for bias applied along the low-conductivity $a$-axis, where bulk transport is strongly suppressed, than for transport along the $b$-axis. However, KPFM alone cannot distinguish enhanced edge conductivity from geometric field redistribution or local conductivity variations. Establishing edge-mediated conduction as the underlying mechanism will therefore require further experimental investigations, complemented by computational modeling and transport simulations to distinguish these possible contributions.
\section{Conclusions}
In conclusion, bias-dependent KPFM provides a direct real-space view of anisotropic charge transport in CrSBr, linking macroscopic transport to local electrical-potential and electric-field distributions. The observed directional dependence and its qualitative agreement with an anisotropic transport model highlight the role of crystal structure in charge propagation. The boundary-associated field redistribution under $a$-axis bias further suggests a possible role of crystal edges in transport along the low-conductivity direction. More broadly, these findings suggest that material anisotropy could be exploited to direct current pathways and enable greater flexibility in device geometry. Future studies combining KPFM with spatially resolved current measurements and controlled edge and interface engineering could clarify these effects and their potential for device design.

\begin{acknowledgments}
\textit{Acknowledgments}- S.S. and M.V. acknowledge the support of the Lumina Quaeruntur fellowship No. LQ200402201 by the Czech Academy of Sciences. A.K. and O.F. thank the Czech Science Foundation (project No. 23-06174K). Z.S. was supported by the ERC CZ program (project LL2101) from the Ministry of Education, Youth, and Sports. This work was also supported by the Ministry of Education, Youth, and Sports of the Czech Republic, Project No. CZ.02.01.01/00/22\_008/0004558, co-funded by the European Union, and the CzechNanoLab Research Infrastructure, supported by the Ministry of Education, Youth, and Sports of the Czech Republic (LM2023051).\\
\\
\textit{Data availability}-- The raw data, analyses, and the related codes underlying this study are available in an online repository at [link to be provided upon acceptance].
\end{acknowledgments}

\nocite{*}
\bibliographystyle{apsrev4-2}
\bibliography{bibliography}
\onecolumngrid
\section*{Table of Contents Graphic}
\begin{figure*}[htbp!]
    \centering
    \includegraphics[width=3.25in]{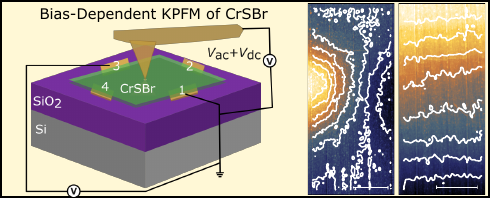}
\end{figure*}



\end{document}



\title{Supplementary Information: Visualizing Anisotropic Charge Transport in CrSBr using Bias-Dependent Kelvin Probe Force Microscopy}

\author{Satyam Sahu$^1$}
\email{satyam.sahu@heu.cas.cz}
\author{Arijit Kayal$^1$}
\email{arijit.kayal@heu.cas.cz}
\author{Bing Wu$^2$}%
\author{Aljoscha Söll$^2$}%
\author{Zdeněk Sofer$^2$}
\author{Matěj Velický$^1$}%
\author{Otakar Frank$^1$}%
\email{otakar.frank@heu.cas.cz}
\affiliation{%
 $^1$Heyrovský Institute of the Czech Academy of Sciences, Dolejškova 2155/3, 182 00, Prague 8, Czech Republic\\$^2$Department of Inorganic Chemistry, University of Chemistry and Technology Prague, Technická 5, 166 28 Prague 6, Czech Republic}%

\date{\today}

\maketitle
\renewcommand{\figurename}{Figure}
\renewcommand{\thefigure}{S\arabic{figure}}
\setcounter{figure}{0}
\renewcommand{\thepage}{S-\arabic{page}}
\setcounter{page}{1}
\renewcommand{\thesection}{S-\arabic{section}}
\setcounter{section}{0}
\renewcommand{\thetable}{S-\arabic{table}}
\setcounter{table}{0}

\onecolumngrid

\section{Raman Spectroscopy of CrSBr}
\begin{figure}[htbp!]
    \centering
    \includegraphics[width=3in]{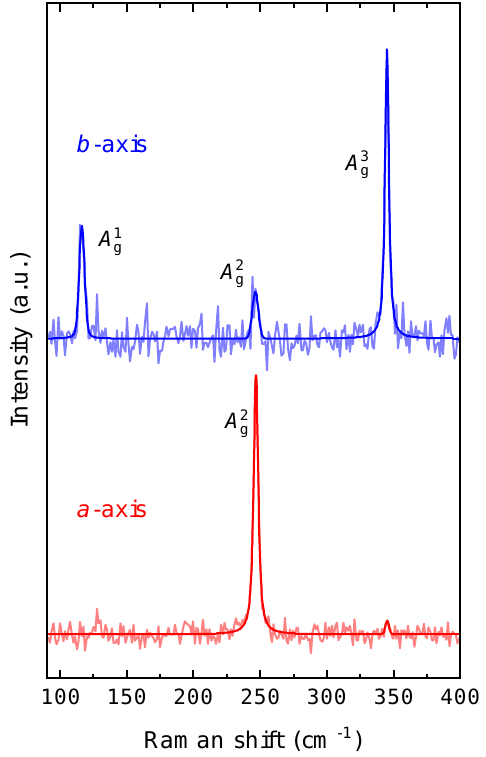}
    \caption{Raman spectra of CrSBr for E$\parallel$$a$-axis and E$\parallel$$b$-axis measured using 2.33 eV excitation.}
    \label{fig:Raman}
\end{figure}

\section{Two-terminal Apparent Field-effect Mobility, Conductivity, and Diffusion Coefficient Estimation}\label{sec:gate-transport}
The apparent charge carrier mobilities along the two axes were calculated within the linear transport regime \cite{Pierret1987, Lundstrom2002} using:
\begin{equation}
    \mu = \dfrac{\left(\dfrac{dI_\textrm{ds}}{dV_\textrm{g}}\right)}{C_\textrm{g} V_\textrm{ds}} \dfrac{L}{W}
\end{equation}
where $C_\textrm{g}$ is a gate dielectric capacitance, given by
\begin{equation}
    C_\textrm{g} = \dfrac{\epsilon_\textrm{r} \epsilon_0}{d_\textrm{hBN}}
\end{equation}
with $\epsilon_0 = 8.854 \times 10^{-14}$ F/cm, $\epsilon_\textrm{r} \textrm{(hBN)}= 3.5$ \cite{Laturia2018}, $d_\textrm{hBN} = 60$~nm, $V_\textrm{ds} = 0.1$~V, L = W = 3 \textmu m.

The 2D sheet conductivity ($\sigma$) was calculated as:
\begin{equation}
    \sigma = n e \mu
\end{equation}

where, $n = 1 \times 10^{13}$ cm$^{-2}$ \cite{Das2025, Liu2026}.
The directional charge carrier diffusion coefficients ($D$) were evaluated via the Einstein-Smoluchowski relation at a room temperature (T = 300 K):
\begin{equation}
    D = \mu \dfrac{k_B T}{e}
\end{equation}
Finally, for anisotropic diffusion, the root-mean-square spreading distances along the principal axes $i$ scale as:
\begin{equation}
    L_i = \sqrt{2D_i t}
\end{equation}
where $t$ is the common elapsed time. For a fixed value of $t$ along the two axes in CrSBr, the relation simplifies to:
\begin{equation}
    \dfrac{L_\textrm{b}}{L_\textrm{a}} = \sqrt{\dfrac{D_\textrm{b}}{D_\textrm{a}}}
\end{equation}
Here, the subscripts $\textrm{a}$ and $\textrm{b}$ refer to the respective principal axis components. Applying this scaling to the measured transverse potential-spreading width gives $L_b^{\textrm{pred,}\sqrt{D}}\approx 4.5$~\textmu m, see main text.\\
Throughout the calculations, it is important to note that the measurements were performed in a two-terminal geometry; therefore, the calculated values do not account for contact resistance or other contributions associated with the measurement configuration.
\begin{figure}[htbp!]
    \centering
    \includegraphics[width=6.5in]{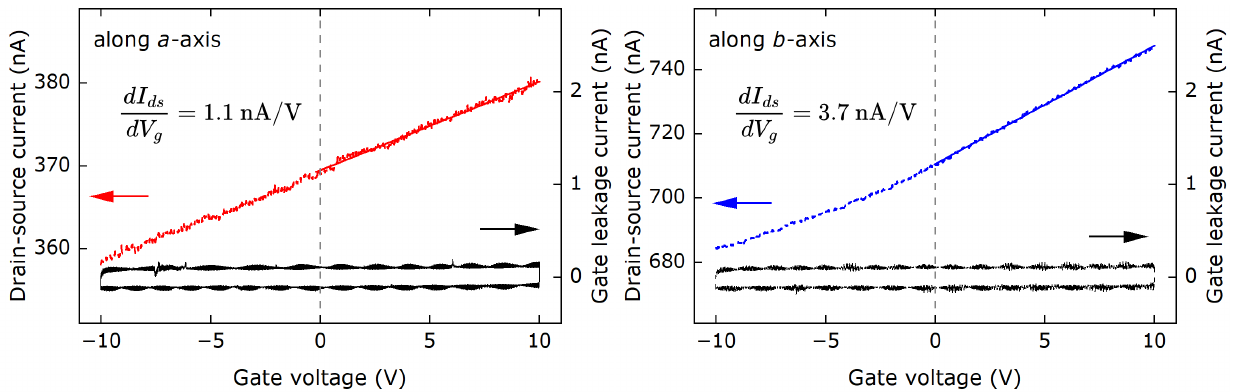}
    \caption{Transport characteristics of CrSBr under applied gate voltage.}
    \label{fig:mobility}
\end{figure}
\renewcommand{\arraystretch}{1.5}
\begin{table}[htbp!]
\centering
\setlength{\tabcolsep}{12pt}
\caption{Directional parameters of CrSBr at room temperature (300 K). (\textit{* square-root diffusivity benchmark estimate.})}
\label{tab:crsbr_transport_metrics}
\begin{tabular}{|c|c|c|}
\hline
\textbf{Transport Parameter} & \textbf{$a$-axis} & \textbf{$b$-axis} \\ \hline
$g_m = \text{d}I_{\textrm{ds}}/\text{d}V_\textrm{g}$ (\text{nA/V}) & $1.043$ & $3.684$\\
\hline
$\mu$ (\text{cm}$^2 $\text{V}$^{-1}$ \text{s}$^{-1}$) & 0.2019 & 0.7133 \\
\hline
$\sigma$ ($\mu$\text{S}) & $0.32$ & $1.15$ \\
\hline
$D$ (\text{cm}$^2$ \text{s}$^{-1}$) & $5.22 \times 10^{-3}$ & $18.4 \times 10^{-3}$ \\
\hline
$L$ ($\mu$\text{m}) & 2.40 \textit{(Measured)} & 4.50 \textit{(Calculated)*} \\
 \hline
\end{tabular}
\end{table}

\section{$I_\textrm{ds}-V_\textrm{ds}$ Output Characteristics}
\vspace{-5cm}
\begin{figure}[H]
    \centering
    \includegraphics[width=6.1in]{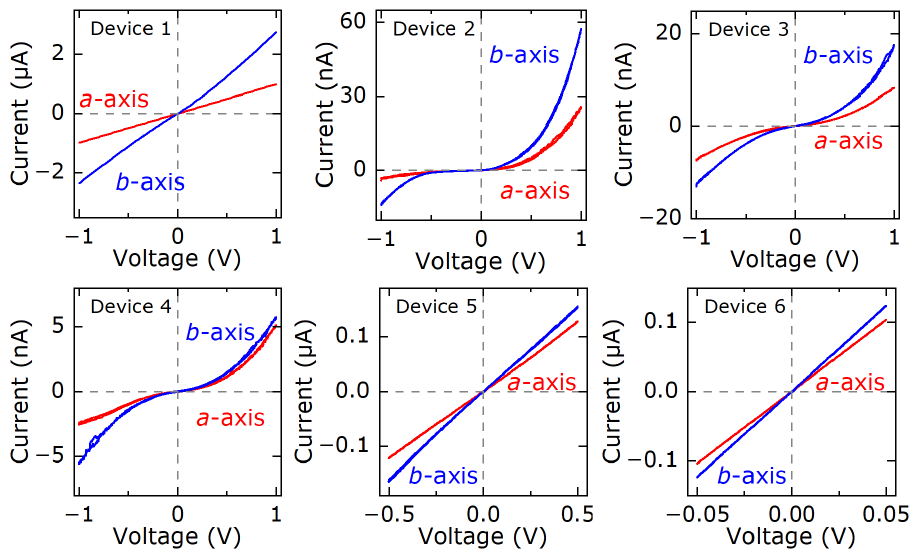}
    \caption{$I_{\mathrm{ds}}$--$V_{\mathrm{ds}}$ curves along $a$- and $b$-axes for different devices.}
    \label{fig:I-V}
\end{figure}

\section{Bias-dependent KPFM for Device 2 (Symmetric Channel Lengths)}
\begin{figure}[H]
    \centering
    \includegraphics[width=6.353in]{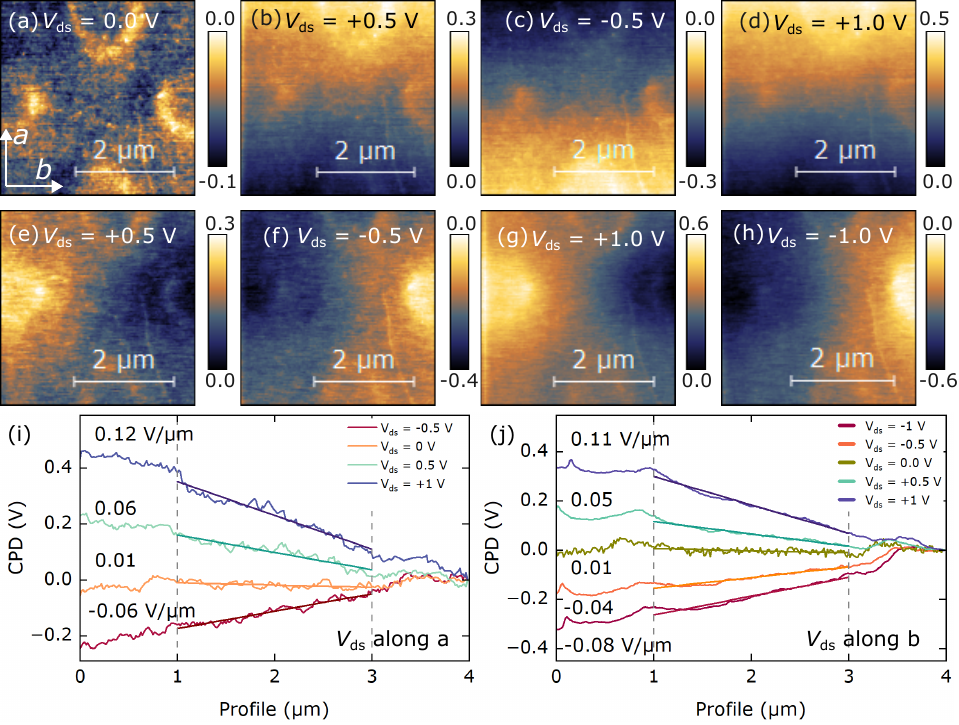}
    \caption{KPFM contact potential maps for a symmetric channel device with (a) zero bias, and bias applied along (b--d) the $a$-axis and (e--h) the $b$-axis. Potential line profiles with geometry-dependent electric field slopes for (i) the low-conductivity $a$-axis and (j) the high-conductivity $b$-axis. The region between the dashed lines is a channel.}
    \label{fig:Device_2}
\end{figure}
\clearpage
\section{Bias-dependent KPFM for Device 1 (Asymmetric Channel Lengths)}
\begin{figure}[htbp!]
    \centering
    \includegraphics[width=6.45in]{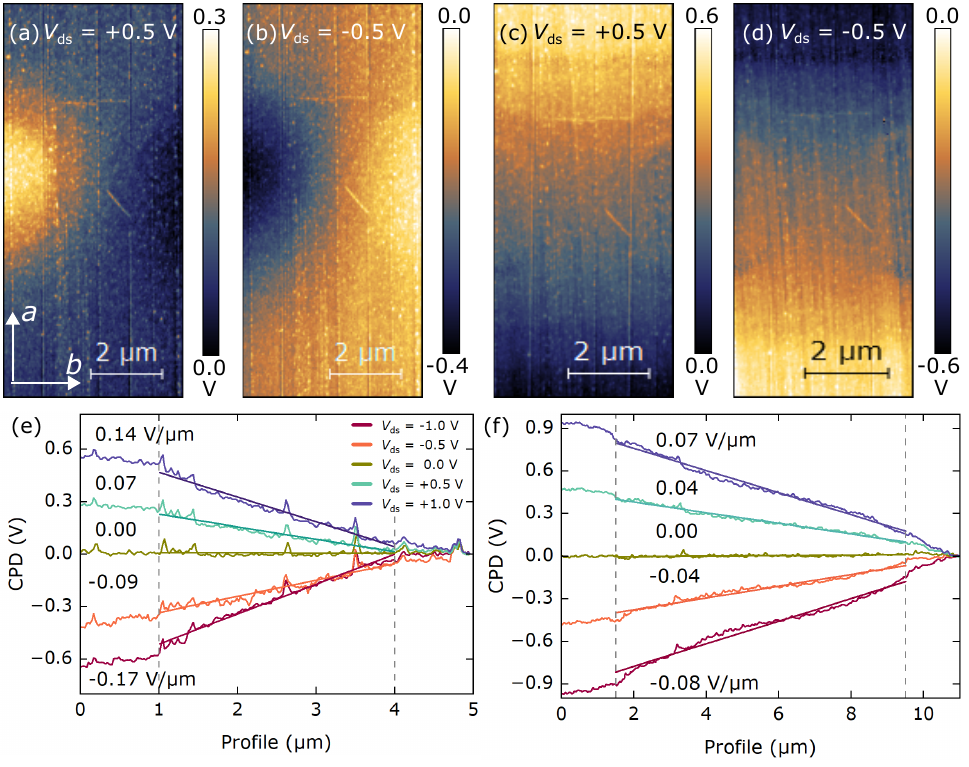}
    \caption{Bias-dependent KPFM contact potential maps under $V_\textrm{ds}$ = $\pm$0.5 V along (a--b) the high-conductivity $b$-axis and (c--d) the low-conductivity $a$-axis. (e--f) CPD line profiles extracted along the $b$- and $a$-axes, respectively.}
    \label{fig:Device_1}
\end{figure}

\clearpage
\section{Equipotential Contours and Local Electric Fields for the Device with Asymmetric Channel Lengths}
\begin{figure}[htbp!]
    \centering
    \includegraphics[width=7.0in]{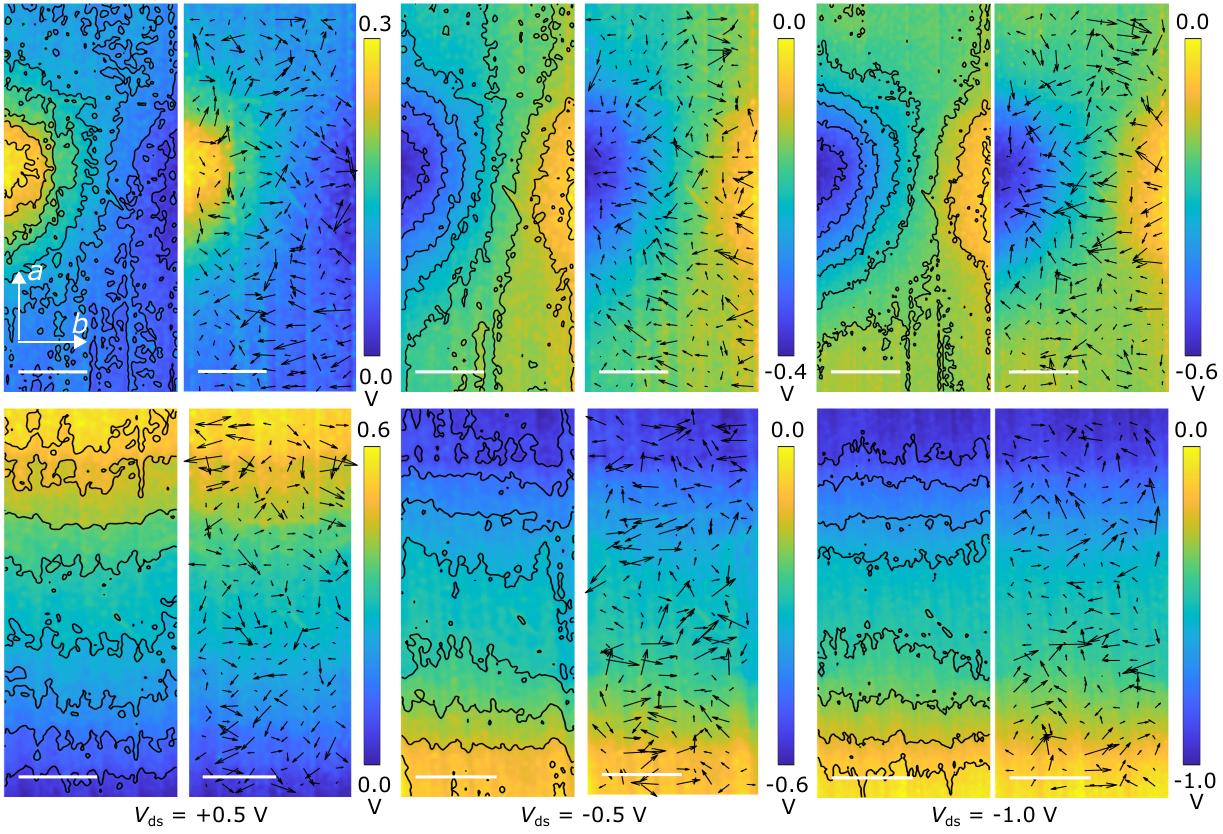}
    \caption{Experimental CPD maps with equipotential contour lines and local electric field for $V_\textrm{ds} = \pm0.5~V$ along the high-conductivity $b$-axis (top row) and the low-conductivity $a$-axis (bottom row). All scale bars: 2 {\textmu}m.}
    \label{fig:EF_ES_asymmetric}
\end{figure}
\clearpage
\section{Equipotential Contours and Local Electric Fields for the Device with Symmetric Channel Lengths}
\begin{figure}[htbp!]
    \centering
    \includegraphics[width=6.5in]{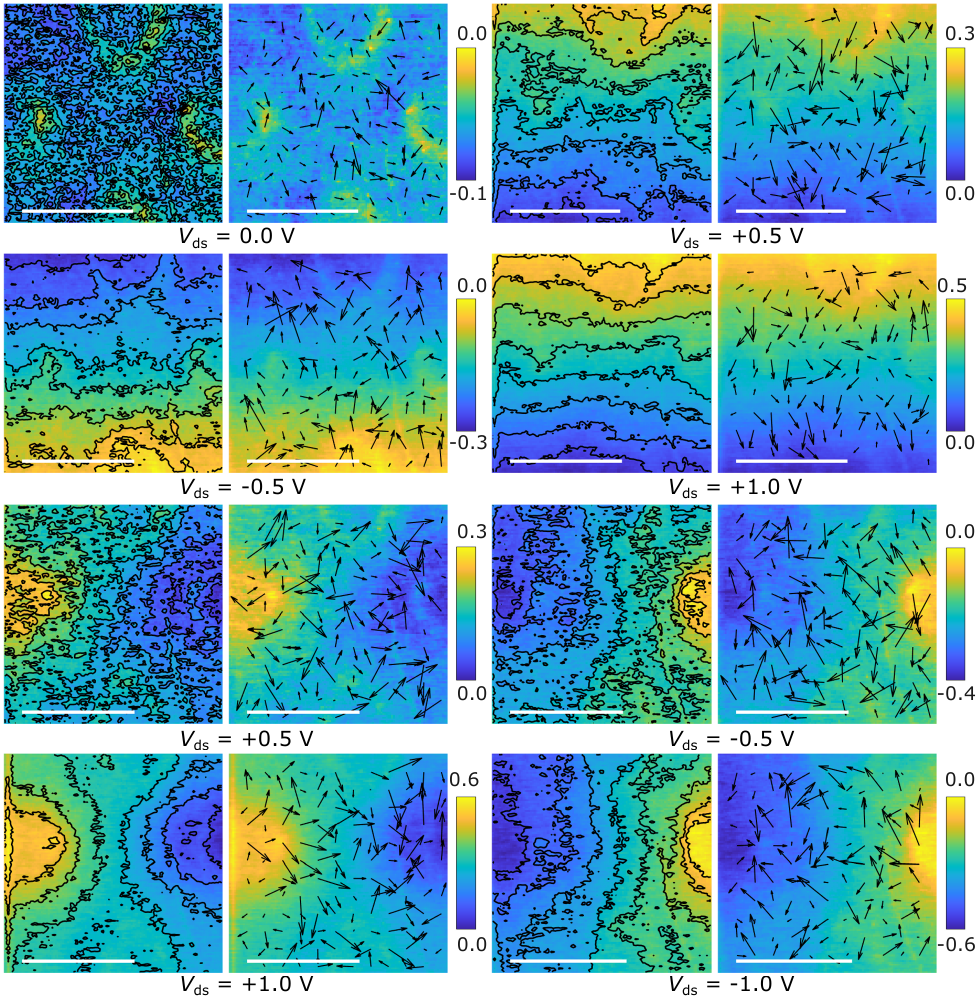}
    \caption{Experimental CPD maps with equipotential contour lines and local electric field for bias along the low-conductivity $a$-axis (top rows) and the high-conductivity $b$-axis (bottom rows). All scale bars: 2 {\textmu}m.}
    \label{fig:EF_ES_symmetric}
\end{figure}

\clearpage
\section{Simulation Methodology and Results}\label{sec:simulation_methodology}
To quantitatively model steady-state carrier transport and investigate non-uniform electrostatic potential landscapes in low-dimensional, structurally anisotropic semiconductors such as multi-layered $\text{CrSBr}$, we establish a continuum transport framework based on the generalized drift-diffusion formulation. Directional variations in atomic orbital overlaps, covalent bonding lengths, and the electronic effective mass tensor ($m^*$) give rise to direction-dependent low-field drift mobilities ($\mu_x, \mu_y$) and spatial diffusion coefficients ($D_x, D_y$) along the distinct in-plane crystallographic orientations. Under an applied bias, the localized current density vectors ($J_x, J_y$) along the principal axes are expressed as \cite{Lundstrom2002}:
\begin{equation}
    J_x = ne\mu_x E_x + eD_x (\partial n/\partial x)
\end{equation}
\begin{equation}
    J_y = ne\mu_y E_y + eD_y (\partial n/\partial y)
\end{equation}
where $n$ represents the local carrier concentration, $e$ is the elementary charge, and $\vec{E} = -\nabla V$ denotes the localized electrostatic driving field vector. Invoking the fundamental directional Einstein relations, the ratio of diffusion flux to drift mobility remains locked to the thermal voltage ($D_x/\mu_x = D_y/\mu_y = k_B T/e$). This thermodynamic coupling guarantees that directional variations in electrical conductivity scale linearly with the diffusion tensor, defining the macroscopic transport anisotropy ratio as $\alpha = \sigma_x/\sigma_y = \mu_x/\mu_y = D_x/D_y$. Under steady-state charge conservation constraints, the system must obey the current continuity condition ($\nabla \cdot \vec{J} = 0$), which yields the governing anisotropic partial differential equation (PDE) for the electrostatic potential landscape $V(x,y)$ \cite{Lundstrom2002}:
\begin{equation}
\alpha \frac{\partial^2 V}{\partial x^2} + \frac{\partial^2 V}{\partial y^2} = 0
\end{equation}
In this framework, the coordinate system ($x,y$) of our device model is assumed to be oriented perfectly parallel with the principal crystallographic axes of the $\text{CrSBr}$ flake. Specifically, the horizontal $x$-axis maps onto the high-conductivity covalent $b$-axis and the vertical $y$-axis maps onto the low-conductivity $a$-axis, treating the crystal as a zero-thickness electrostatic membrane decoupled from the out-of-plane $z$-axis. This purely diagonal principal-term solver is physically validated by the quasi-1D electronic wire model established for $\text{CrSBr}$ in the literature \cite{Wu2022}, where a complete absence of the Hall effect confirms that charge carriers are tightly confined within independent, isolated covalent chains and cross-talk is restricted to random, incoherent interchain quantum tunneling events. Because coherent transverse deflection and macroscopic shear currents are structurally suppressed by the material's atomic layout, a principal diagonal framework remains physically accurate regardless of contact placement asymmetry.

To numerically solve this governing equation, the continuous domain is discretized onto an orthogonal spatial mesh grid of size $N_y \times N_x$ with uniform step sizes $\Delta x = L_x/(N_x - 1)$ and $\Delta y = L_y/(N_y - 1)$. Applying second-order central finite-difference stencils to approximate the spatial derivatives converts the differential equation into an algebraic relaxation update stencil calculated at every internal node $(i,j)$:
\begin{equation}
V_{i,j}^{(k+1)} = \frac{c_x \left( V_{i+1,j}^{(k)} + V_{i-1,j}^{(k)} \right) + c_y(j) \left( V_{i,j+1}^{(k)} + V_{i,j-1}^{(k)} \right)}{2 \left( c_x + c_y(j) \right)}
\end{equation}
where $c_x = \alpha / \Delta x^2$ represents the horizontal transport stencil weight, and $c_y(j) = 1/ \Delta y^2$ controls the localized vertical transport weight. Unbiased internal nodes are initialized with a neutral midpoint seed vector ($V_{(i,j)} = 0.5\text{ V}$) to optimize convergence rates and prevent numerical shock across the boundaries. The matrix is updated iteratively for up to $800$ successive sweeps until the global residual satisfies a convergence criteria of $\sum |V^{(k+1)} - V^{(k)}| < 10^{-6}\text{ V}$. Following convergence, the driving field vectors are mapped directly using $\vec{E} = -\nabla V$, to overlay magnitude-scaled arrows natively over the landscape.\par
To establish the boundary conditions for the numerical simulations, fixed-potential (Dirichlet) conditions were applied at the source and drain contacts, with the corresponding grid nodes held at \(V_{\mathrm{bias}}\) and 0 V, respectively. At open flake edges, zero-normal-field (Neumann) conditions, \(\partial V/\partial n=0\), were imposed to prevent current flow across the flake–vacuum boundary. Floating contacts were treated as equipotential metallic regions, with a uniform potential assigned across each contact footprint and adjusted to ensure zero net current into the floating contact.\par
To replicate the characteristic background disorder inherent to exfoliated 2D flake landscapes, the simulation injects a multi-featured microscopic defect profile. The coordinates $(x_p, y_p)$ of these disorder centers are distributed randomly across the crystal domain using a standard uniform random number generator to achieve a spatially uncorrelated topography. The peak energetic amplitudes of these features follow a normal (Gaussian) distribution derived via random numbers generated from a standard normal distribution, $w_p \sim \mathcal{N}(0,1)$, ensuring a balanced representation of both electropositive impurity landscapes and electronegative carrier traps. The local profile of each individual potential puddle is modeled using an anisotropic two-dimensional Gaussian envelope function. 
\clearpage

\begin{figure}[htbp!]
    \centering
    \includegraphics[width=6.6in]{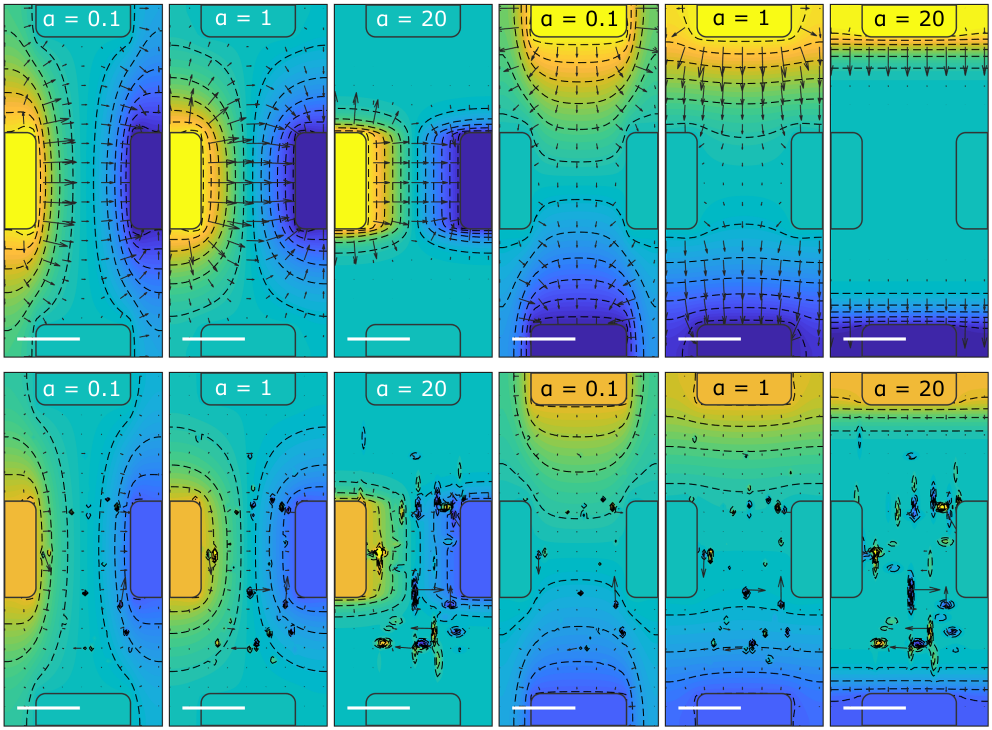}
    \caption{Vectorized numerical simulations modeling potential contours and electric field lines for bias along the $b$- and $a$-axes for different anisotropy ratios ($\alpha$ = 0.1, 1, and 20) (top row) and the corresponding contours and electric field lines for the point defects (bottom row) for a device having asymmetric channel lengths.}
    \label{fig:simulation_asymmetric}
\end{figure}

\begin{figure}[htbp!]
    \centering
    \includegraphics[width=6.6in]{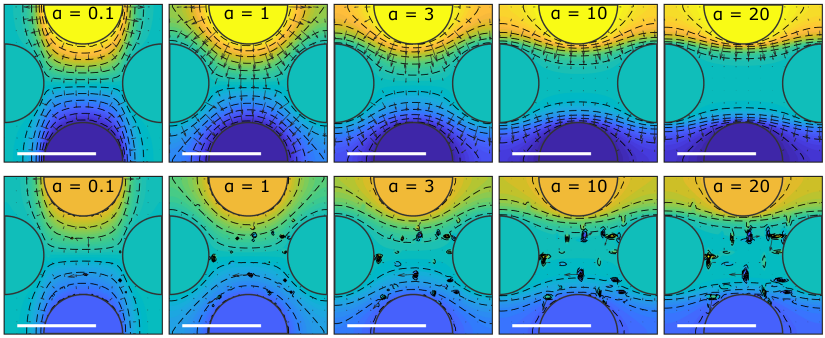}
    \caption{Vectorized numerical simulations modeling potential contours and electric field lines for bias along the $a$-axis for different anisotropy ratios ($\alpha$ = 0.1, 1, 3, 10, and 20) (top row) and the corresponding contours and electric field lines for the point defects (bottom row) for a device having symmetric channel lengths.}
    \label{fig:simulation_symmetric_long_axis}
\end{figure}

\begin{figure}[htbp!]
    \centering
    \includegraphics[width=6.6in]{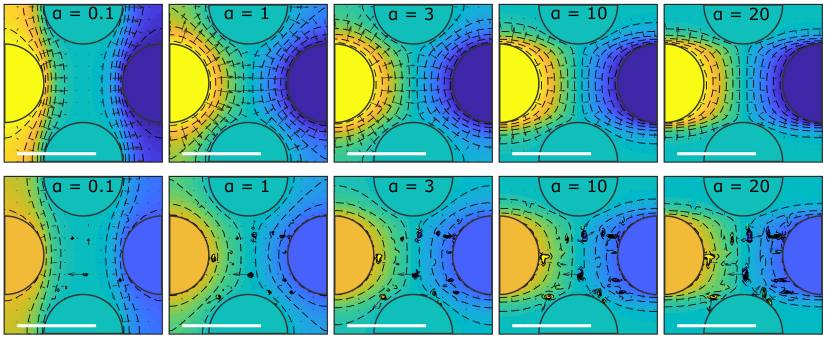}
    \caption{Vectorized numerical simulations modeling potential contours and electric field lines for bias along the $b$-axis for different anisotropy ratios ($\alpha$ = 0.1, 1, 3, 10, and 20) (top row) and the corresponding contours and electric field lines for the point defects (bottom row) for a device having symmetric channel lengths.}
    \label{fig:simulation_symmetric_short_axis}
\end{figure}

\begin{figure}[htbp!]
    \centering
    \includegraphics[width=7.0in]{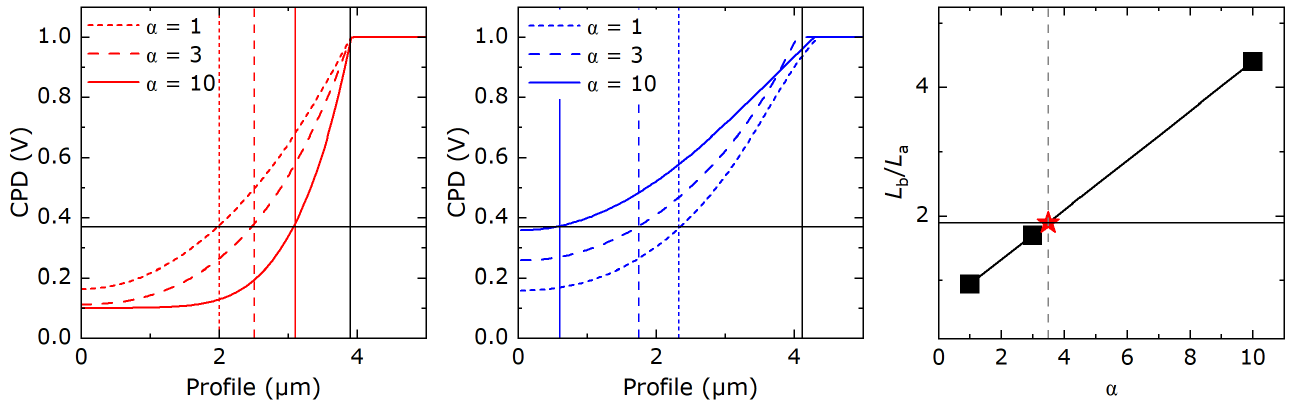}
    \caption{Transverse line profiles for bias along the $b$- and $a$-axes from the simulations for $\alpha = 1, 3,$ and $10$ and the corresponding $L_\textrm{b}/L_\textrm{a}$ ratios. The red star symbol corresponds to the expected $L_\textrm{b}/L_\textrm{a}$ for the experimental anisotropy.}
    \label{fig:transverse}
\end{figure}
\nocite{*}
\bibliographystyle{apsrev4-2}
\bibliography{bibliography_SI}

